\documentclass[sn-standardnature]{sn-jnl}

\jyear{2021}%

\theoremstyle{thmstyleone}%
\theoremstyle{thmstyletwo}%

\theoremstyle{thmstylethree}%

\begin{document}

\title[How to write your astronomy paper]{How to write and develop your astronomy research paper}


\author*[1,2]{\fnm{Johan H.} \sur{Knapen}}\email{johan.knapen@iac.es}
\equalcont{These authors contributed equally to this work.}

\author[3]{\fnm{Nushkia} \sur{Chamba}}\email{nushkia.chamba@astro.su.se}

\author[4]{\fnm{Diane} \sur{Black}}\email{d.black2021@outlook.com}
\equalcont{These authors contributed equally to this work.}

\affil*[1]{\orgname{Instituto de Astrof\'\i sica de Canarias}, \orgaddress{\street{Calle V\'\i a L\'actea S/N}, \city{La Laguna}, \postcode{E-38205}, \state{}\country{Spain}}}

\affil[2]{\orgdiv{Departamento de Astrof\'\i sica}, \orgname{Universidad de La Laguna}, \orgaddress{\city{La Laguna}, \postcode{E-38200}, \state{}\country{Spain}}}

\affil[3]{\orgdiv{The Oskar Klein Centre, Department of Astronomy}, \orgname{Stockholm University}, \orgaddress{\street{AlbaNova}, \city{Stockholm}, \postcode{SE-10691}, \state{}\country{Sweden}}}

\affil[4]{\orgdiv{Free-lance English coach}, \orgname{}\orgaddress{\street{}\city{Groningen}, \postcode{}\state{}\country{The Netherlands}}}


\abstract{Writing is a vital component of a modern career in scientific research. But how to write correctly and effectively is often not included in the training that young astronomers receive from their supervisors and departments. We offer a step-by-step guide to tackle this deficiency, published as a set of two papers. In the first, we addressed how to plan and outline your paper and decide where to publish. In the current second paper, we describe the various sections that constitute a typical research paper in astronomy, sharing best practice for the most efficient use of each of them. We also discuss a selection of issues that often cause trouble to writers, from sentence to paragraph structure, the `writing mechanics' used to develop a manuscript. Our two-part guide is aimed primarily at master's and PhD level students who are presented with the daunting task of writing their first scientific paper, but more senior researchers or writing instructors may well find the ideas presented here useful.}




\maketitle

\section{Introduction}\label{sec1}

This is the second paper in a mini-series aimed at providing basic guidelines and sharing best practice for scientific writing, especially to beginning writers of refereed papers in astronomy. In Paper I (Chamba, Knapen \& Black 2022) we discussed how to prepare to write your paper and where to submit it. We also described how you can organise your paper so the message comes across clearly and effectively. We provided references to other resources available, including an eclectic mix of refereed papers, editorials and author instructions, books, and online publications. 

In this Paper II, we discuss the organisation of a professional scientific paper in the general field of astronomy, and how to make the various sections of a paper effective in bringing your message across to the readers. We emphasise structure and clarity as tools to help you deliver the narrative of your work and paper. We close by providing background and best practice regarding the mechanics of writing a paper. This includes aspects like paragraph and sentence structure, and a number of specific aspects of written English which complicate the lives of beginning and experienced writers alike, in particular but by means exclusively, when English is not their native language. 

\section{Step by Step Guide to Writing a Scientific Manuscript}

\subsection{Title: Clear, Informative, Short}

The title is the main marketing tool for your paper---it is hard to get right but you need a good one to catch a potential reader's attention. Ideally, it captures the main message of the paper, the `narrative', or story, in around 15 words. The title needs to be accurate and needs to match the abstract and the rest of the paper. Keep the title as short as possible. Use simple words, no jargon, no abbreviations, no new concepts. Include all important key words (search engines and other indexing tools will later on allow others to locate your paper). Avoid vague titles. Avoid starting with `On...' or `Towards...' (instead, say what you've done!). Avoid jokes, references to modern culture which may not age well, or titles that may offend. If possible, package the main conclusion of your paper into it. 

As an exercise, you can take a recent issue of a major journal and consider what you think of the various paper titles. Do they inspire you to read the abstract? Or put you off reading further? Then compare your own draft title(s) with what you have seen.

\subsection{Abstract}

The abstract sets the scene, identifies the problem, outlines how you will solve that problem, and then describes what you have done, how, and what you have concluded (ideally coming back to `setting the scene'). Often, it ends with an outlook---what is the next step after what we have learnt here. In the abstract, you try to tell the whole story of your paper in perhaps 150 words.  The editors at the time of A\&A described the importance of an abstract \cite{Bertout2005} `as a filter for deciding what articles are worth taking the time to read in detail'. They then suggest that authors should provide an abstract that `conveys the essential elements of the article to the reader: its objective, the methods used to reach it, and the results obtained' (\cite{Bertout2005}). The conclusions of the work should also be described.  

A useful exercise is to compare abstracts from different papers, for instance when you make a literature review (Sect.~2.1). Which ones call your attention, and invite you to read the paper? How many fail to set the stage and start by telling you what they did, rather than why? If you do not have the time to read the whole paper, do you think the abstracts will allow you to vaguely remember the results later on? Can you recognize the elements described above as essential for an abstract?

\subsection{Figures and Tables}

It is often useful to develop an idea very early on about which figures and tables will need to be included in the paper. You will need to find the balance between including all the key ones, without including too many. Series of figures and tables, as well as long tables, can be published electronically only, and/or in an Appendix. As usual, a journal's {\href{https://journals.aas.org/graphics-guide/}{instructions}} for {\href{https://www.nature.com/documents/NRJs-guide-to-preparing-final-artwork.pdf}{authors}} will inform you how to produce your tables and figures. Good tips on how to produce high-quality tables and figures are given by \cite{Sterken2011b} and \cite{Chen2022}.

Figures need to be easy to understand and of high quality. Try to make them as vector graphics (pdf is vectorial, jpeg is not) so they can be scaled up. Always deliver figures of the highest possible quality to the publisher, even if you use smaller versions for other purposes (e.g. ArXiV or talks). Include all lines, points, labels, etc. that you need, but omit any that are not strictly necessary. Explain all elements of the figure in the caption. Do not make your figures too complicated: you want them to catch the reader's attention. Interactive or animated figures can help you convey lots of, or complicated, information (e.g. Fig. 1 in \cite{Peterken2019}). Try to make your figures tell the story of the paper: some people will look at the figures in your paper, or see them in a conference presentation, without reading your carefully worded text (sometimes even without reading the caption). Actively looking for the key message in other papers in title, abstract, and figure captions can help you try to act like the audience of your own paper when writing it! 

In terms of presentation, use a colour scheme that can be interpreted by all, including those with colour vision deficiency (`colour blindness', affecting up to 10\% of your colleagues). Simple but effective tips include using both line/point style and colour to distinguish different elements (never only colour), and using a colour scheme that is appropriate for colour-blind people. Many \href{https://jfly.uni-koeln.de/color/index.html}{resources} are \href{https://bconnelly.net/posts/creating_colorblind-friendly_figures/}{available} online.
 
We recommended to use the same format for all your figures, at least in terms of typeface, text size, colour coding, etc. You can end each caption with a one-sentence conclusion of what the figure shows, or make the figure title into one. That way, if a reader scans the figures of your paper, or someone tweets your figure, the context and implications directly come with it. 

\subsection{Building up your Paper}
\label{subsec:building}

The typical structure of a research paper in our field is the following: Introduction, Methods/Sample/Data, Results, Discussion, Conclusion (IMRaD). Try not to mix these sections. Keep them logical and as short as you can. Use sub-sections and paragraphs to separate the text into `bite-sized chunks' that can be digested easily by a reader. These sections collectively build your paper into the story, or `narrative' that you defined in your outline.

Keep an overall `hourglass' shape of the paper in mind: it starts broadly, with the `big picture' in the introduction, then narrows down into the details of methods and results, before broadening again at the end, when the discussion and conclusion sections place the new findings in a wider context.

\subsection{The Introduction}
\label{subsec:introwriting}

The Introduction has several roles, which you can use in this order to draft it.

\begin{enumerate}
\item	Present the problem in a wide context. The `Big Picture'. Example `topic' sentences (Sect.~3.1) could start with `The Sun is...' or `In modern cosmology, ...'
\item	Narrow the focus by introducing the specific sub-topic you will be describing, e.g. `magnetic fields around sunspots...' or `galaxy mergers ...'
\item	Point out the gaps in the existing knowledge which you want to fill in with your work, e.g. `however, the exact role of... is unclear'.
\item	Now present what you did in your study, and define the exact problem tackled. `The aim of the current study is...'
\item	Finally, describe what you have done, `We therefore observed/modelled...'
\item	In long papers, the Introduction can end with a preview of what follows. `This paper presents...' or `we present our sample in Sect 2...'
\end{enumerate}

A good introduction is concise, focusses on the main issues, avoids repetition and includes references to relevant work. This allows the reader to place the work in context, and to find key background papers if they want to know more. You can also introduce other relevant aspects such as a specific telescope, instrument, simulation, or survey, or a particular object. The introduction should present all the background information that is relied upon in the rest of the paper. Ideally new `background' should not appear in the discussion section or conclusions.

Keep the introduction short and reasonable. Unless you write a full review paper, you do not need to refer to every paper ever written on the topic. Use `e.g.' and then give the most important or relevant ones. You could cite a review paper but this risks not properly crediting those who did the original work (see \cite{Chen2022}). To find the key papers in your field (which may have been published before you were born), use citation statistics to find the most-cited papers in a certain field, then analyse why they are so highly cited. The key papers will soon become clear.

Make sure you cite modern work, including from the current year. This shows that your research is relevant and timely. On the other hand, do not {\it only} include modern papers. In any case read and check all papers you cite to make sure they in fact support your claim and do not assume that previous authors have done this.  

Consider for what statements you need a reference. Basically, for textbook-level basic knowledge, no; specific statements or results, yes. Do not be afraid to cite your own relevant papers, or those of your friends. Do not overdo it either, papers that only quote own work are bound to have less impact. 

\subsection{Sample, Data, Methods}

This is probably both the most technical section of your paper, and the easiest to write (often also the first section that researchers write). In an observational paper, for instance, you describe what data you used, how you obtained and reduced them, and what analysis methods you used. Do not describe the results---that comes in the next section. Justify, however, why your data or methods are needed to meet the goals of your paper.

The main aims of this section are to allow others to (1) judge how you reached your results, and how reliable, good, novel, etc. your data and methods are, and (2) reproduce what you did. While point 1 is generally straightforward to assess, point 2 reflects an open issue and ongoing debate in astronomy. There are many solutions to make a scientific paper reproducible, each with their own caveats (see for example \cite{akhlaghi2021}, \cite{Kuttel2021}). It is important to include all critical details in the paper. Too often when we refer to older papers it is impossible to reproduce the methods used in detail because the data or methods were not described well enough, are unusable, or even not publicly available anymore. Do not hide any caveats of your methods, and be honest---identify and address either how you have solved these issues or why they do not affect your results (if possible, in a quantifiable way). On the other hand, do not describe everything you ever did, and ignore issues that do not affect the main line of the paper. 


In some journals (such as Nature or Science) the Methods section is placed at the end of the paper, or put into a separate online section called the `Supplementary Information' or `Supplementary Materials'. In other journals you can place non-critical or background information in an appendix. In either case, make sure you reference the background material in the main paper. Be honest, and if there is a significant problem or doubt, do not bury it in an appendix or in the supplementary materials. 

Keep the most technical parts in specific subsections or paragraphs, so a reader can skip them if they want. Finally, we recommend using the past active tense: `we used...' (not `we use...' or `we have used...'), but more important than that is to be consistent: use whatever you choose throughout the Section.

\subsection{Results}

In astronomical professional writing, this is where you describe what you have found. It is good idea to foreshadow which are the main points that will be discussed later, but there is no need to discuss all the implications yet, or to make detailed comparisons with other work, because that will come in the Discussion section. 

The results section is illustrated with figures. Select those first, define (or remind yourself of) your narrative, and use a logical structure by employing sub-sections and paragraphs. Section and sub-section titles allow the reader to scan the paper quickly. Keep  (sub-)section titles short but include the key words or even the key message you want the reader to pick up even if they just scan them and do not read the actual text.

An effective writing trick is to start each section and sub-section with a sentence saying why you looked at the particular aspect you are about to describe. Compare these two ways of starting: (i) {\it In Figure 3 we plot our data for all the stars in our sample}; and (ii) {\it To demonstrate how parameter x relates to y, we show the results for all the sample stars in Figure 3}. The second approach clearly leads the reader to look at Figure 3 while foreshadowing what they can expect.

\subsection{Discussion}
\label{subsec:discwriting}

This is where we come back to the hourglass shape of the paper (Sect.~\ref{subsec:building}). The previous sections, methods and results, were very specific and probably very technical. In the discussion, you place your results in the context of other work, or compare them to theory, modelling, or observations. The other main aim of the discussion section (which is really `critical discussion') is to describe the limitations of the results you have just presented. 

Once again, use sub-sections with well-chosen titles to break up your discussion into manageable parts, choose your paragraphs strategically, and use topic sentences (Sect. 3.1) to set the scene. Then include final sentences or paragraphs in each section or subsection to summarise the main point.

Some key aspects you can include in the discussion section, broadening up the `hourglass' as you go down the list, are the following.

\begin{itemize}
\item[1.]	Repeat why you carried out this study.
\item[2.]	Return to the main question you posed early on in the abstract, and answer it.
\item[3.]	Explain how your answer is supported by your data, model, figures, derivations, results.
\item[4.]	Discuss how your results, and your answer, relate to other work in the literature. Say how they are supported, how they support other work, but also where they disagree.
\item[5.]	Give your view on why there is disagreement, and/or discuss any alternative explanations, but without distracting the reader from the main message of your paper.
\item[6.]	Broaden up to describe wider implications, applications, and recommendations.
\item[7.]	Finally, add what future work or observations can provide more clarity or advancement.
\end{itemize}

\subsection{Conclusions}

Always include your conclusions. Many readers do not have time to read the whole paper, but may skip to the conclusions section to get a quick overview of what you have done and found. If you do not make a separate Conclusion section, then at least call the previous one `Discussion and Conclusions' (and make sure you explicitly add the conclusions in the text). 

Start the conclusions section by summarizing very briefly the why, how and what of your paper. Repeat key points like what the aim of the study was, the sample size, with what telescope you observed the objects or which pipelines or codes you ran, and then make a concise and logical summary of the main results and how you interpret them. You may want to refer back to your key figures in the conclusions section, to tempt those readers who read only this section to also look at your artwork.

Finish up with a strong concluding sentence. Do not just say something bland like `more research is needed' but try to summarise your whole paper in one final sentence. Something like `our results show how galaxy metallicities confirm...'.

\subsection{Acknowledgements, References, Appendices}

Acknowledgements are important. This is both in a formal sense, for instance because your funding agency will insist that you acknowledge their financial support, and in an informal one, because you want to record your appreciation of the help you have received from others---colleagues who helped you with specific aspects of the work, or friends or family who are not colleagues but who for some reason you feel deserve to be thanked publicly for their role in the work you have managed to complete and publish. The Acknowledgements section is also the adequate place to thank those who have provided any public data or code you rely on, and often telescope archives and software/model creators will ask for specific statements to be included.

References will need to be included in the form prescribed in the instructions for authors of the particular journal you have chosen. Ensure that the references are complete and correct because their use is to allow others to find the papers. Do not save on citations---our colleagues are our referees too, and everybody likes to see their own work referred to by others. Consider whether you can, or need to, add software or other products which are not strictly research papers to your formal reference list. Citations are often needed by the relevant authors or developers to secure future funding, or to show that their work has been useful. Where references are not needed or possible, then add the software or other products to the acknowledgements section. 

Finally, use appendices to include material that would make the paper too long, or that is necessary but too technical to include in the main body. Sometimes you can also move additional tests or explorations to the appendices. As appendices are almost always published electronically, length restrictions tend to be either absent or less of an issue. 

\section{Writing Mechanics for Manuscript Development}

Section 2 detailed the IMRaD scheme for writing your manuscript. In this section, we provide guidelines on various structural aspects of writing in the English language, the `mechanics' of writing paragraphs and grammatically correct sentences for linkage. In the Supplementary Information, we share some more detailed hints on how to use correct punctuation with relative clauses (e.g., {\it the section which Diane wrote} versus {\it the section, which Diane wrote}, and on when and how to use infinitives and gerunds as nouns (e.g., {\it she advises reading it} versus {\it she advises to read it}).

\subsection{Paragraphs}

Inexperienced writers may have problems writing well-structured paragraphs. For example, they may find themselves writing wandering paragraphs without any clear point. The reader tries to follow, but starts to wonder `where is this paragraph going? What is the point the author is trying to make?' Often the problem is the lack of a `topic sentence'.

Well-structured paragraphs begin by expressing the main idea in the first sentence (topic sentence), then develop that main idea with facts, examples, or analysis. This structure is common in academic writing.  In other types of writing, for example novels, the structure may be looser.  Furthermore, paragraphs may follow a different structure in other languages, for example, in Chinese, where the main idea is expressed at the end of the paragraph.  We address paragraph structure in the English language with an example. 

What is the point of the following paragraph? 
{\it In 2018 Brown measured x.  Similarly Jones published y. Recently Smith reported z.}
Perhaps the writer intended to show that there has been a lot of research on galaxy Q. In this case, a good topic sentence could clarify the significance of these facts. 
{\it Recent research has focussed on the galaxy Q. In 2018 Brown measured x.  Similarly Jones published y. Recently Smith reported z.}

Another common problem for beginning writers is trying to cram too much into one paragraph. A paragraph must contain just enough information to explain your ideas to the reader, but avoid unnecessary detail. Paragraphs should be neither too long nor too short. Nowadays, paragraphs in academic writing tend to contain three to seven sentences, though the length is somewhat journal-dependent. What do you think is wrong with the following paragraph?

{\it We used several approaches to measure x. First we did a. Then we did b and c. After measuring x, we also measured y.}
This paragraph begins by describing the measurement of x, but then gets off track by unexpectedly mentioning the measurement of y. There are two possible ways to fix this problem. Either change the topic sentence to include the measurement of y ({\it we used several approaches to measure x and y}) or put the measurement of y into a paragraph of its own. In any case, as a general rule of thumb, avoid tremendously long and discouraging-looking paragraphs that go half a page or more. Split the long ones up into two smaller ideas, each with its own topic sentence.

A topic sentence may be very general, for example {\it a galaxy is a group of stars}. This very general sentence could lead to a description of the types of galaxies. Such general sentences are common at the beginning of an introduction. A topic sentence in a methods section could be something like {\it to measure x, we followed several steps}. This topic sentence very obviously leads to a list of steps, probably using words such as {\it first, next, finally}. Topic sentences in a results section often refer to a figure or table. {\it Table 3 shows the measurements of xxxx}. This sentence would then lead to a systematic presentation of each result. Paragraphs do not need a special sentence to end, but sometimes in a results or discussion section, it is helpful to give a conclusion, for example {\it these results suggest that xxxx}. Another way to end is with a transition sentence, for example, {\it the next Section describes xxxx}. 

A paragraph can end when the necessary information has been given, or it can provide a summation or linkage to the next idea. 

\subsection{Sentence Linkage}

Beginning writers, particularly non-anglophones, often write bumpy text that does not seem to flow. This problem can be fixed by linking sentences to each other. In the English language the beginning of a sentence always provides context and linkage, generally by repeating a word or concept that has already been mentioned, while the end of the sentence presents the new information (in some languages, such as Persian, the structure is just the opposite: the new information comes at the beginning of the sentence, while the old, familiar, linking information comes at the end).  Here are several examples showing good linkage.

{\it Galaxy W was recently discovered by Brown.  This galaxy...} Renaming the noun {\it galaxy} links the sentences and leads probably to a description of the galaxy.

{\it Galaxy B has been well characterized. In 2018 Brown reported..., while a year later Jones published...} In this case, `well characterized' suggests research, which is listed in the following sentences, so this structure leads to a good flow. The use of the dates provides a bit of extra context, showing that this is recent research.

{\it There are currently three missions studying the comet. First...  Second... Third...} Here, the specific mention of `three missions' (or perhaps `several missions') leads the reader to expect the list that follows.

Another technique to improve linkage and flow is the use of `sign-post' words that tell the reader the relationship of the ideas.  To list examples, you can use phrases like `for example,' `for instance,' and `such as.' To show chronological order, use dates like `in 2012' or words like `first,' `second,' `next,' and `finally.' To show cause and effect, use words or phrases like `consequently' or `as a result.' To suggest similarity, use words like `similarly' or `likewise.' To show contrast, use phrases like `in contrast', `on the other hand' or `however.'

\subsection{Tips for Nice Sentences}

Creating a flow between the sentences that make up a paragraph requires that each individual sentence is well structured. Some tips for writing nice sentences are as follows.

{\bf Make sentences just the right length}. Sentences in academic writing range on average between 18 and 32 words, with some longer or shorter. Shorter sentences can provide a nice antidote if the previous sentence was rather long.  But sentences can get too long and too complicated by the stringing together of clauses, especially those that begin with `which' or `that.' Aim to make these longer complex sentences represent no more than one out of three or four in your paper.

{\bf Stick to a subject-verb-object structure}.  This structure puts the verb right after the subject, where it belongs.  So not `characteristics such as a, b, and c were measured,' but much better `we measured a, b, and c.' 

{\bf Prefer active voice}.  Active voice leads to more lively writing, as demonstrated by the example in the previous paragraph.  Using active voice does not mean beginning every sentence with `we.' Inanimate objects can also be the subject: {\it this paper presents...; the next section explains...; the results show...; this model provides...}.

{\bf Use verb tenses correctly}.  The main verb tenses used in academic writing are Present, Present Perfect, and Past.  

Present tense for  
\begin{enumerate}
    
\item[1.] facts and definitions \\
 	Ex: {\it These giant low surface brightness galaxies form an interesting class of objects.}	 

\item[2.]	referring to the paper or figures or tables in the paper \\
 	Ex: {\it Fig. 1 shows.../ X is shown in Fig. 1/this paper presents...} 

\end{enumerate}

Past tense for things you did in the past, or that somebody else did in the past. \\
 	Ex: {\it For each bin, we calculated the surface brightness.}  \\
 	Ex: {\it The European Space Agency launched the {\it Gaia} satellite at the end of 2013.} 

Present perfect  
\begin{enumerate}
    
\item[1.] to announce work done in the `recent' past with emphasis on the current relevance.\\ 
 	Ex: {\it We have explored the nature of the disc truncations in two edge-on nearby Milky Way-like galaxies.}\\ 
 	Ex: {\it Algorithms have been proposed along this line.}

\item[2.]	for a process that began in the past and continues up to now.\\
 	Ex: {\it Over the last twenty years, there has been intensive research into...}

\end{enumerate}

{\bf Keep subject and verb close together.}

Not: A number of galaxies, some with spirals arms and others with both spiral arms and a prominent bar, were observed.

But: We observed a number of galaxies, some ... etc.

{\bf Use appropriate verbs.} People often mix the action we take, as observers, or researchers, with what an object like a star or a galaxy does. 

An example: {\it Galaxies concentrate a lot of dust in their central regions}. (Not necessarily, we see that the dust is concentrated there but did the galaxy do it? Write {\it A lot of dust is concentrated in the central regions of galaxies}.).

\subsection{Place Modifiers Carefully} 

A modifier is a word or phrase which gives additional information about another word or phrase. 

Look at this example: {\it orbiting at x km above the earth, we could see the satellite}.  Are you really orbiting above the earth, or is it the satellite? Verbal phrases, like {\it orbiting at x km above the earth} should be placed directly before or directly after the word they modify.  So you can correctly write {\it orbiting at x km above the earth, the satellite could be seen with the naked eye} or {\it we could see the satellite orbiting the earth at x km above the earth}.

What about {\it last weekend when I was camping, I looked out of my tent and could see the satellite in my pyjamas}? What is that satellite doing in your pyjamas?  Prepositional phrases like {\it in my pyjamas} usually work best right after the noun they modify. In some cases, the sentences really need a bit of rewrite.  We could put the part about the pyjamas into a separate sentence or drop it, and tell about {\it the satellite in the Western sky}.

\subsection{Avoid Ambiguity with {\it This}}

A common mistake is referring to something with `this', without defining what it refers to. Such as here:

{\it In many galaxies, secular evolution leads to the formation of pseudo-bulges and also to other changes in their morphology. This can be observed in both optical and radio imaging.} What can be observed? The secular evolution? The formation of pseudo-bulges? Or the changes? To solve this problem, you write {\it this evolution can be observed}.

\subsection{Tricky Plurals}

While most plurals in English are formed by adding `s' or `es,' science uses some irregular plurals. `Data' is usually used as a plural, with the singular form being `datum' or `a piece or item of data.' In singular/plural pairs, here are a few tricky plurals: analysis/analyses, spectrum/spectra, phenomenon/phenomena, criterion/ criteria, matrix/matrices, index/indeces.

\subsection{Use of the Articles a/an/the}

Articles ({\it the} dog, {\it a} galaxy) can be extremely tricky for many non-native speakers because many
or even {\href{https://wals.info/chapter/37}{most}} people in the world do not use articles at all in their first language, or do not use
them like in English! It is thus very common in their drafts to see `the' and other articles where an English native speaker
would not place one, or not to see an article where an anglophone would place one.

You should also note that for native English speakers (or for speakers of languages
which treat articles in the same way as English, such as Dutch or German and to a
lesser extent French and Spanish) the use of articles is often completely intuitive and
crystal-clear, and they may not understand how others can possibly have a problem with articles.

Here are the basic rules.
    
1. Does the reader know which one or ones you are talking about?  Then use `the.'
The reader might know which one you are talking about for the following reasons: 

\begin{itemize}
    
\item[a.]	Known because you just introduced it. \\
{\it There was an association between a and c.  The association remained robust after...}
\item[b.]	Known because you mention something related. \\
{\it ...galaxy x.  The structure of the galaxy is...} [every galaxy has a structure] 
\item[c.]	Known because you define it in the sentence.\\ 
{\it The causes of this rotation are...}
\item[d.]	Known because you name it. \\
{\it We used data from the SIMBAD database.}  [You define the database by giving the name, SIMBAD] 
\item[e.]	There is only one.  The Milky Way. The Sun. However, do not use `the' for the names of people.  {\it I sent the data to Peter} (not `the Peter'). Do not add `the' to names of countries, but be aware that `the' forms part of the name of some countries (the United States, the Netherlands). {\it The conference was in Spain; next year it will be in the Netherlands}.

\end{itemize}

2. If rule 1 does not apply and the noun is  singular, use `a' or `an.' `A' is used before a consonant sound: a galaxy, a star, a DVD. `An' is used before a vowel sound: an element, an hour, an X-ray emitter.	

3. If rule 1 does not apply and the noun is plural or noncountable, do not use any article (some languages do not differentiate between singular and plural nouns, which leads to confusion with articles and matching up the verbs.  For example, in Indonesian, the structure is something like `one book is brown; two book is blue.' If this situation is the case in your mother language, be careful to make nouns singular and plural, and to match the verb to the noun in English, so you have `one book is brown; two books are blue.') Generally speaking, abstract terms used in a broad, general sense are noncountable and are not introduced with `the'. 

Examples with abstract nouns used in a general sense:\\
{\it You need more information about the university program}. (`Information' is used in a general sense.)\\
{\it Phrasal verbs can be used in speech, but not in academic writing}. (Both `speech' and `academic writing' are used in a general sense.)

But when using these abstract terms in a restricted and specific sense, rule 1 applies, and `the' is used.
 
Examples with abstract nouns used in a restricted, specific sense:\\
{\it The information in that folder is not correct}. (Specific information)\\
{\it The speech you hear on the street in New York is different from the speech you hear in London}. (Specific examples of speech)\\
{\it The writing in this paper is excellent}. (Specific example of writing)

4. Some confusion may arise because some nouns can be either countable or noncountable, depending on their usage. To make matters worse, some nouns are countable in English but not in other languages, and vice versa. 

Examples of nouns used as countable:\\
{\it We gave the students some exercises to help them practice the use of articles}. (countable, plural `exercises')\\
{\it I was so busy in the lab yesterday! I did 72 analyses}. (`analyses' refers to analysed samples)

Examples with the same nouns used as noncountable:\\
{\it You should not sit in front of the computer all day; you should get some exercise}. (`Exercise' in general---maybe walking or cycling. This is quite a different meaning than the written exercises to practice the use of articles in the example above)\\
{\it He entered the data into the computer for analysis}. (In this case, `analysis' is used as a general abstract noun, like `information' in the example above.)

For those whose native language does not use articles, practice the rules (you can find exercises on the internet), or analyse some of the sentences in this paper. Can you identify why an article is used or not? You will not perfect your use of articles overnight, but with some regular attention to the issue, you can improve a lot within a year or two. For those who are anglophones, remember that articles are very difficult for your nonanglophone colleagues. Be helpful, not critical.  After all, is your Indonesian/Russian/Persian perfect?

\subsection{Adverbs}

An adverb describes a verb, an adjective, or another adverb. It tells us how, where, when, how much, how often, etc. Adverbs of time (such as never, often, sometimes, also) usually do not cause trouble, but the adverbs that end in -ly can be confusing.

They are formed:

\begin{itemize}

\item For most adjectives, add -ly: wonderful---wonderfully
\item If an adjective ends in y, change y to i and add -ly: easy---easily
\item Ending in -able or -idle or -ible, change -e to -y: horrible---horribly
\item Ending in c, add -ally: automatic---automatically
\item Some adverbs are irregular and do not end in -ly: good---well
\item And some do not change at all: fast---fast.

\end{itemize}

In fact, academic writing does not use a lot of adverbs to modify adjectives or adverbs, except in the special case of past participles used as adjectives.  Examples are {\it previously reported data, a recently launched satellite, a carefully written report}.

\subsection{Amount vs Number}

Use `number' for countable objects, and `amount' for measured quantities.

Correct: {\it The number of stars of a given mass...}

Correct: {\it The total amount of fuel needed to launch the satellite...} (or else {\it the number of litres of fuel}, since litres are countable).

Correct: {\it The amount of time needed for the project...} (or else {\it the number of months needed for the project}, since months are countable).

\subsection{Avoid Phrasal Verbs in Formal English}

English uses many phrasal verbs, such as `break down' or `talk about,' but these are informal. They can be used in speech but not in formal writing.  In almost all cases, there is a single, more formal word that can replace the phrasal verb, for example, `destroy' or `decompose' are more formal than `break down,' and `discuss' is more formal than `talk about.' The more formal versions are derived from a Latin root and often resemble words in French or Spanish. Other examples are using the formal `investigate' instead of `look into' or the formal `continue' rather than `go on.'

\section{Concluding Remarks}

This paper is the second part of a self-help guide to writing papers in astronomy for students and beginning writers. Often, the expectation is that students will learn how to write papers from their mentors or supervisors but not all mentors are the best of writers or are trained, or even willing, to teach students how to write. We hope our guide provides students with a clear framework to begin writing their first paper, or perfect their subsequent ones.

Without a doubt, if you continue to practice and pay attention to good planning and writing, it will become easier to produce well-written papers. We also remind graduate programme coordinators that writing in astronomy is an essential skill for the future success of young researchers. We believe that the writing experience of students would be greatly enhanced if programmes included courses in academic writing as part of their early scientific education or graduate research training.

In Paper I, we provided guidelines on how to plan your paper, where to submit it, and how to organise your paper. In this Paper II, we discuss the organisation of the paper and the mechanics of writing and polishing text. We concentrate on scientific papers to be submitted to refereed professional journals in astronomy.  Based on our experience, we believe that every researcher can learn to write clearly, competently and even gracefully. We provide these guidelines based on our experience as writers and teachers, but take these guidelines as just that---only guidelines which will help you find your own way to becoming a good writer. Your reward will be when a colleague compliments you on your beautifully written paper.

\backmatter

\bmhead{Acknowledgments}

We thank S\'ebastien Comer\'on, Sim\'on D\'\i az-Garc\'\i a, Erik \& Laura Knapen Almeida, Cristina Mart\'\i nez-Lombilla, Rainer Sch\"odel and Aaron Watkins for comments on an earlier version of these notes. NC also thanks T. Emil Rivera-Thorsen, Matteo Pompermaier and Chris Usher for interesting discussions. Part of this paper is based on a scientific writing course delivered by JHK to mostly MSc and PhD students in Ethiopia and Rwanda. He wishes to thank Professors Mirjana Povi\'c and Pheneas Nkundabakura for organising those courses, and the students for participating.

\section*{Declarations}

{\bf Funding} J.H.K. acknowledges financial support from the State Research Agency (AEI-MCINN) of the Spanish Ministry of Science and Innovation under the grant `The structure and evolution of galaxies and their central regions' with reference PID2019-105602GB-I00/10.13039/501100011033, from the ACIISI, Consejer\'{i}a de Econom\'{i}a, Conocimiento y Empleo del Gobierno de Canarias and the European Regional Development Fund (ERDF) under grant with reference PROID2021010044, and from IAC project P/300724, financed by the Ministry of Science and Innovation, through the State Budget and by the Canary Islands Department of Economy, Knowledge and Employment, through the Regional Budget of the Autonomous Community. NC acknowledges support from the research project grant `Understanding the Dynamic Universe' funded by the Knut and Alice Wallenberg Foundation under Dnr KAW 2018.0067.

\bmhead{Authors' contributions} 

All authors developed the ideas in this manuscript together. J.H.K. is the primary author, D.B. wrote most of Sect. 3, and all authors contributed to editing of the manuscript. The development of this guide was inspired by the scientific writing courses which D.B. has been giving for years at the University of Groningen and, at N.C.'s initiative, also at the Instituto de Astrof\'\i sica de Canarias, Tenerife in 2019.

\bmhead{Competing Interests} 

The authors declare no competing interests.

\bibliography{NatAst_P2_v2}

\newpage

\section*{Supplementary Information}

\subsection*{A. Correct Punctuation with Relative Clauses}

First we need to define a few grammatical terms.  A {\it clause} is a group of words that contains a subject and a verb.  So a clause might be a whole sentence that can stand alone (an independent clause). Or a clause might not be able to stand alone (a dependent clause).  Here we are concerned with a particular class of dependent clauses, those that begin with {\it which, who}, or {\it that}. This type of clause is known as a relative clause, since the pronoun ({\it which, who}, or {\it that}) relates back to a previously mentioned noun. Here is an example of a sentence containing a relative clause: {\it Prof. Smith, who discovered the galaxy, won the Nobel prize}. In this case, {\it who} refers to Prof. Smith.

The words {\it which} and {\it who} can be used in two ways: to give defining information or extra, nondefining information. 

\underline{Defining information} is used to specify which person or thing you are talking about.

Example: {\it The professor who has a red shirt discovered a new galaxy}. In this case, there are several professors in the room, and your friend is pointing out which one you should look at. Thus the clause `who has a red shirt' defines which professor your friend is talking about. Note that there are no commas around the defining clause.

When talking about people, we usually use the pronoun {\it who}, but for things, it is possible to use {\it that} or {\it which} to introduce defining information. 

Example: {\it The galaxy which was discovered last week is amazing}. Note that here the clause `which was discovered last week' defines which galaxy we are talking about, so there are no commas.  In this case, we could also use {\it that} instead of {\it which}: {\it the galaxy that was discovered last week is amazing}. Some writers prefer to use {\it that} for defining information and save {\it which} for nondefining information (see below), but in fact both these words are in common use for the defining clauses.

\underline{Nondefining information} gives extra information about the person or thing, but it is already clear which person or thing you are talking about. 

Example: {\it Prof. Smith, who discovered the new galaxy, will give the next lecture}. In this case, we know exactly which professor your friend is talking about: Prof. Smith.  The clause `who discovered the new galaxy' gives nondefining information, so in fact this clause could be put into a separate sentence like this: {\it Prof. Smith will give the next lecture.  She discovered the new galaxy}. Note that in this case, the commas around the defining clause indicate that this is extra, nondefining information that you could put some place else. 

Note that the pronoun {\it that} is never used to give nondefining information.

Who cares?  Why does it matter whether I use commas or not? Well, sometimes, putting in commas or leaving them out changes the meaning of the sentence. 

Example without commas (defining relative clause): {\it The model which I developed fits the data}. In this case there are many models under discussion, but the particular one that I developed fits the data.

Example with commas (nondefining relative clause): {\it The model, which I developed, explains the data}. In this case, there is only one model under discussion, it explains the data, and by the way, I developed it. The information about who developed the model could be put in a separate sentence: {\it I developed a model. The model explains the data.}

In summary, if you need the information in the relative clause to specify who or what you are talking about, then do not use commas around the clause.  If the information could just as well be put into another sentence, then do use commas around the clause. 

\subsection*{B. Infinitives and Gerunds as Nouns}\label{Appendix2}

Confusingly, English uses both infinitives and gerunds as nouns. The infinitive is the `to' form, like `to eat,' and the gerund is the noun form that ends in `ing,' like `eating.' In some cases, either the infinitive or the gerund can be used:  {\it we began to investigate the star} or {\it we began investigating the star}. But depending on the main verb, it may be necessary to use the infinitive or the gerund. {\it We wanted to investigate the star} because `want' requires the infinitive.  {\it We considered writing a paper} because `consider' requires the gerund.  Some verbs that are commonly used in academic writing can be used in several different ways. For example, {\it the supervisor advised taking more observations} because `advise' requires the gerund. It is also correct to use the structure {\it the supervisor advised me to take more observations}; the infinitive is required when the person receiving the advice is mentioned. In case of doubt about the correct usage, check a dictionary such as The Longman {\href{https://www.ldoceonline.com/}{Dictionary}} of Contemporary English.

\end{document}